\documentclass[aps,pra,twocolumn]{revtex4}

\usepackage{graphicx}
\DeclareGraphicsExtensions{.eps}

\usepackage{xcolor}
\usepackage{amsmath}
\usepackage{qcircuit}
\usepackage{tikz}
\usetikzlibrary{automata}
\usepackage{blindtext, rotating}
\usepackage{braket}
\usepackage{multirow}
\usepackage{tabularx}

\begin{document}

\title{Optimizing quantum phase estimation for the simulation of Hamiltonian eigenstates}

\author{P. M. Q. Cruz$^{1,2,}$\footnote{pedro.cruz@quantalab.uminho.pt}, G. Catarina$^{1}$, R. Gautier$^{1,3}$, J. Fern\'andez-Rossier$^
{1,}$\footnote{On leave from Departamento de F\'isica Aplicada, Universidad de Alicante, 03690 San Vicente del Raspeig, Spain.}
}

\affiliation{$^1$QuantaLab, International Iberian Nanotechnology Laboratory (INL), Av. Mestre Jos\'e Veiga, 4715-330 Braga, Portugal \\
$^2$Departamento de F\'isica e Astronomia, Faculdade de Ci\^encias da Universidade do Porto, 4169-007 Porto, Portugal \\ 
$^3$CentraleSup\'elec, Universit\'e Paris-Saclay, 91190 Gif-Sur-Yvette, France \medskip{}}

\date{\today}

\begin{abstract}

We revisit quantum phase estimation algorithms for the purpose of obtaining the energy levels of many-body Hamiltonians and pay particular attention to the statistical analysis of their outputs. We introduce the mean phase direction of the parent distribution associated with eigenstate inputs as a new post-processing tool. By connecting it with the unknown phase, we find that if used as its direct estimator, it exceeds the accuracy of the standard majority rule using one less bit of resolution, making evident that it can also be inverted to provide unbiased estimation. Moreover, we show how to directly use this quantity to accurately find the energy levels when the initialized state is an eigenstate of the simulated propagator during the whole time evolution, which allows for shallower algorithms. We then use IBM Q hardware to carry out the digital quantum simulation of three toy models: a two-level system, a two-spin Ising model and a two-site Hubbard model at half-filling. Methodologies are provided to implement Trotterization and reduce the variability of results in noisy intermediate scale quantum computers.

\end{abstract}

\maketitle

\section{Introduction} 

The computational resources required to model a quantum system in a classical computer scale exponentially with the number of degrees of freedom. This is known as the exponential wall problem \cite{kohn99}. As a result, complete numerical solutions of the general many-body problem, where reduction schemes for the Hilbert space of the system are impossible or unknown, can only be achieved for very small systems. This precludes the simulation of interesting molecules and their chemical reactions with the so-called chemical accuracy \cite{langhoff2012}. Finding a viable universal approach to solve the many-body problem, going around the exponential wall, would enable tremendous progress in scientific areas such as condensed matter physics and quantum chemistry.

In this context, Feynman put forward the notion that quantum simulation should be used to circumvent the exponential wall \cite{feynman82}, even before quantum computers were envisioned \cite{Deutsch85}. Once the concept of the gate based universal quantum computer was established, quantum algorithms were proposed \cite{lloyd97, lloyd99} to tackle the many-body problem. This approach to the simulation of quantum systems is referred to as Digital Quantum Simulation (DQS) \cite{preskill2018}, to distinguish it from analog quantum simulators \cite{georgescu14}. In the context of quantum chemistry, DQS enables the computation of molecular energies \cite{aspuru2005, lanyon10, omalley2016} or other physical quantities \cite{wecker15}. In general, DQS could be used to tackle the many-body problem both in condensed matter \cite{aspuru2005, wecker15, cervera2018, chiesa2019} and quantum field \cite{preskill2018qft} theories.

There are now several quantum and quantum-classical algorithms that would permit to address the many-body problem if sufficiently powerful quantum computers were available. Outstanding examples include the simulation of time evolution \cite{lanyon2011} and the variational quantum eigensolver (VQE) \cite{peruzzo2014, kandala2017, gard2019}. The latter is a heuristic approach to find the groundstate energy by classically optimizing the parameters of a quantum ansatz. This hybrid algorithm has been more recently proposed as a better suited methodology for Noisy Intermediate Scale Quantum (NISQ) \cite{preskill2018} computing devices. However, the former strategy is still worth pursuing for being an exact method to approximate quantum dynamics which, when combined with the Phase Estimation Algorithm (PEA) \cite{cleve98, Nielsen} or the Iterative PEA (IPEA) \cite{griffiths96, Dobsicek07}, permits to obtain the full energy spectrum of many-body Hamiltonians \cite{aspuru2005, lanyon10, omalley2016}, unlike the VQE. 

In the last few years, quantum computing hardware has experienced a qualitative leap with the advent of cloud-based quantum computing platforms. Motivated both by the availability of quantum hardware and the potential of quantum computing to tackle the many-body problem, we explore the implementation of Quantum Phase Estimation (QPE) based DQS algorithms in the NISQ computers of IBM \cite{ibmq}. At the time of writing, the IBM Q platform permits remote access to quantum computers with 5, 16, 20 and 53 superconducting qubits, and is being used by dozens of research groups worldwide to explore a broad set of applications \cite{devitt2016, alsina2016, fedortchenko2016, garcia17, wootton2017, coles18, cervera2018, murta2020berry}.

Phase estimation procedures are very important to determine the eigenvalues of a given unitary operator \cite{kitaev1996quantum, cleve98, Nielsen}. Their applications span several areas including factorization \cite{shor1994algorithms}, sensing \cite{higgins2009demonstrating}, gate calibration \cite{kimmel2015robust} and, relevant for this work, quantum simulation \cite{aspuru2005, lanyon10, omalley2016,obrien2019}. There are two main strategies for algorithmic QPE. The first makes use of the gate expensive inverse quantum Fourier transform (QFT) \cite{cleve98} and, in an ideal quantum computer, could work with a single shot readout. The second uses much shallower circuits, such as the one proposed by Kitaev \cite{kitaev1996quantum} or the so-called iterative PEA \cite{griffiths96, Dobsicek07}, but requires multiple readouts and classical processing.

There is a large body of work dedicated to optimize both variants of QPE algorithms. In \cite{svore2013faster}, an extension of Kitaev's approach \cite{kitaev1996quantum} was studied, offering a logarithmic reduction in the number of necessary measurements to estimate the phase with exponential accuracy. More recently, a post-processing method based on classical time-series analysis of QPE measurements was shown numerically \cite{obrien2019} to be capable of determining multiple eigenvalues simultaneously when initializing general quantum states. Another approach to estimate several eigenvalues simultaneously based on time-series methods is introduced in \cite{somma2019quantum}. Here, we go back to the original QFT-based QPE and introduce a novel methodology to optimize its use for NISQ computers. Our approach is based on a new estimator which can be employed to find eigenvalues of both hermitian and unitary operators. We focus on the former use case and examine the determination of the energy levels of three simple model Hamiltonians: 1) a two-level system, 2) a two-spin Ising model and 3) the two-site Hubbard model at half-filling. Our results have implications in a broader context, given that QPE algorithms are a central subroutine in quantum computing \cite{Nielsen}.

The successful implementation of the PEA-based quantum simulation requires a relatively modest number of qubits, but a rather large number of quantum gates. The number of qubits is determined by the number of single-particle states, plus a small overhead to readout the results. Given that the largest exact classical computations of fermionic Hamiltonians cannot deal with more than 30 single-particle states, 50 qubits would be enough to achieve quantum supremacy in the context of DQS. However, current state-of-the-art hardware is far from the depth required to make PEA-based DQS work beyond the simple models considered in this work.

The rest of this paper is organized as follows. In section \ref{sec:DQS} we summarize the main steps of DQS based on QPE algorithms. In section \ref{sec:CPP} we review the PEA and the IPEA and present our approach for the classical post-processing of the quantum measurements that permits to improve the DQS results. In section \ref{HAM} we introduce two simple spin model Hamiltonians to which the DQS is carried out, as well as the gate implementation of their unitary evolution operators. In section \ref{sec:spinres} we present the quantum computation results obtained for the spin models. In sections \ref{sec:hub} and \ref{sec:hubres} we introduce the two-site Hubbard model at half-filling and the quantum simulation results, respectively. In section \ref{dc} we wrap up the main results and conclusions. Additional technical details are provided in the appendices.

\section{DQS via Quantum Phase Estimation \label{sec:DQS}}

In this section, we summarize the theory of DQS based on QPE algorithms that permit to obtain the eigenvalues of model Hamiltonians. In particular, we illustrate the main concepts using the PEA. A prerequisite is to be able to encode the quantum states of the Hamiltonian in qubit states. In the case of spin $S=1/2$ models, this is straightforward. In the case of fermions, there are canonical transformations such as the Jordan-Wigner transformation that provide a mapping between second quantization fermion operators and spin operators. 

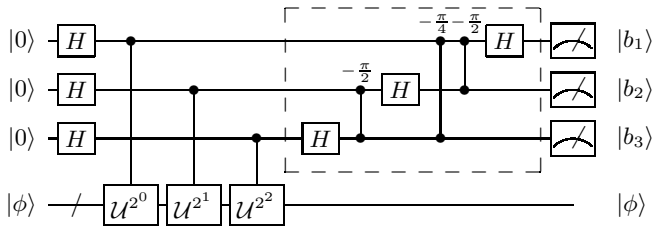
\begin{figure}[hbt]
\begin{centering}
\begin{tabular}{c}
\Qcircuit @C=0.38em @R=0.6em {
   \lstick{\left|0\right\rangle} & \gate{H} & \ctrl{4} & \qw & \qw & \qw & \qw & \qw & \qw & \qw & \qw & \qw & \ctrl{2}_{-\frac{\pi}{4}} & \qw & \ctrl{1}_{\;\;\;-\frac{\pi}{2}} & \qw & \gate{H} & \qw & \qw & \meter & \rstick{\left|b_1\right\rangle} \\
   \lstick{\left|0\right\rangle} & \gate{H} & \qw & \ctrl{3} & \qw & \qw & \qw & \qw & \ctrl{1}_{\;-\frac{\pi}{2}} & \qw & \gate{H} & \qw & \qw & \qw & \ctrl{-1} & \qw & \qw & \qw & \qw & \meter  & \rstick{\left|b_2\right\rangle} \\
   \lstick{\left|0\right\rangle} & \gate{H} & \qw & \qw & \ctrl{2} & \qw & \gate{H} & \qw & \ctrl{-1} & \qw & \qw & \qw & \ctrl{-2} & \qw & \qw & \qw & \qw & \qw & \qw & \meter  & \rstick{\left|b_3\right\rangle} \\
  & & & & & & & & & & & & & & & & & & & \\
  \lstick{\ket{\phi}} & \qw {/} & \gate{{\cal U}^{2^0}} & \gate{{\cal U}^{2^1}} & \gate{{\cal U}^{2^2}} & \qw & \qw & \qw & \qw & \qw & \qw & \qw & \qw & \qw & \qw & \qw & \qw & \qw & \qw & \qw & \rstick{\ket{\phi}}
\gategroup{1}{7}{3}{17}{1.4em}{--}}
\end{tabular}
\par\end{centering}
\caption{Quantum circuit that implements the Phase Estimation Algorithm with 3-bit resolution. $\mathcal{U}$ is a general unitary operator that acts on a multi-qubit eigenstate $\ket{\phi}$ as $\mathcal{U} \ket{\phi} = e^{i 2 \pi \phi} \ket{\phi}$. In the context of quantum simulation, $\mathcal{U}$ is the time evolution propagator of the simulated Hamiltonian. The boxed operations illustrate the permuted inverse quantum Fourier transform. The output is a rational approximation to $\phi$ given by $0.b_1 b_2 b_3$.}
\label{fig:PEA}
\end{figure}

The PEA is schematically shown in Fig.~{\ref{fig:PEA}}. The circuit uses two registers: the {\em phase register}, on top, has $R$ qubits ($R=3$ in the case of the diagram) to encode a phase approximation with one of the $2^R$ possible readout states; the lower line(s) correspond to the multi-qubit {\em simulation register} where a relevant many-body state is loaded. The procedure has the following steps:
\begin{enumerate}

\item {\em Initialization}. An eigenstate $\ket{\phi}$ of the target Hamiltonian $\mathcal{H}$, with energy $\varepsilon$, is prepared in the lower register of the diagram. When the state is known, there is an algorithm to carry out this task \cite{shende06}. Except for simple model Hamiltonians, eigenstates are in general unknown and therefore the input is a linear combination of eigenstates. The consequences are discussed in the next section.

\item {\em Unitary evolution}. This subroutine entails the representation of (powers of) the unitary evolution operator ${\cal U} = e^{-i {\cal H} {\tau/\hbar}}$ into gates that act on $\ket{\phi}$ as ${\cal U} \ket{\phi} = e^{i 2 \pi \phi} \ket{\phi}$ and are controlled by the top register qubits. Most often, $\mathcal{H}$ is a sum of non-commuting terms and a Trotter-Suzuki approximation is required to implement ${\cal U}$. It is convenient to consider dimensionless variables for energy, $\varepsilon\rightarrow\varepsilon\,\varepsilon_0$, and time, $\tau\rightarrow\tau\hbar/\varepsilon_0$, so that the dynamical argument of the propagator is given by $2\pi \phi(\tau)=-\varepsilon\tau$. Importantly, in DQS, $\tau$ is a simulated parameter rather than actual computational time.

\item {\em Phase estimation}.
The core of the PEA has two stages. First, the controlled ${\cal U}^{2^{k-1}}(\tau)$ operations kick out the fractional digits \cite{Nielsen} of the phase to the top register. Second, the inverse Fourier transform permits to obtain them through a measurement in the computational basis, so that readout gives access to the phase $\phi(\tau)$. Repeating the procedure for several values of the control time parameter $\tau$, the eigenvalue $\varepsilon$ is obtained from the variation of the phase with $\tau$,
\begin{equation}
\varepsilon=-2\pi \frac{\Delta \phi}{\Delta \tau}.
\label{eq:eig_est}
\end{equation}

\item {\em Post-processing}. Two main obstacles arise in the previous program. First, even for an ideal quantum computer, both the fact that the initial state is in general a linear superposition of $\ket{\phi_n}$ eigenstates and the fact that $\phi_n$ cannot always be expressed as a binary fraction decomposition, lead to quantum dispersion of the readouts. Second, noise and readout errors have to be dealt with when using NISQ computers. Thus, the statistical post-processing of QPE algorithms becomes very important and is the subject of the next section.
\end{enumerate}

\section{Post-processing \label{sec:CPP}}

This section is devoted to the statistical methods that we have implemented to enhance the phase estimation procedure described in the previous section. Before presenting the methods, we briefly review some technical aspects of QPE algorithms.

We explore two different QPE algorithms to obtain Hamiltonian eigenvalues: the so-called PEA \cite{Nielsen} and IPEA \cite{griffiths96, Dobsicek07}. Both address the following problem: given an unitary operator $\cal{U}$, and one of its eigenstates $\ket{\phi}$, such that ${\cal U} \ket{\phi} = e^{i2 \pi \phi}\ket{\phi}$, PEA and IPEA yield an estimate of the phase $\phi \bmod 1$.

In order to approximate a real positive number smaller than 1, a binary fraction representation of $\phi$ with $R$ bits is used,
\begin{equation}
0.\phi = b_1 2^{-1}+b_2 2^{-2}+ b_3 2^{-3}+ \cdots + b_{R} 2^{-R}.
\label{eq:binrepr}
\end{equation}
For a given value of $R$, this representation defines a grid of rational numbers separated by $\delta \phi=2^{-R}$. For instance, in the case of $R=3$  (see circuit of Fig.~\ref{fig:PEA}), there are 8 possible readout states, $\ket{000}$, $\ket{001}$, $\ket{010}$, $\ket{011}$, $\ket{100}$, $\ket{101}$, $\ket{110}$, and $\ket{111}$, that represent $0$, $0.125$, $0.250$, $0.375$, $0.5$, $0.625$, $0.75$, and $0.825$, respectively.

The Hadamard gates acting on the phase register prepare all the qubits in the state $\frac{1}{\sqrt{2}} \left(|0\rangle+|1\rangle\right)$. The controlled ${\cal U}^{2^{k-1}}$ operators kick-back \cite{cleve98,Nielsen} a phase into the $k^{\mathrm{th}}$ qubit of the phase register $\frac{1}{\sqrt{2}} \ket{0}+e^{i 2\pi {2^{k-1}}\phi} \ket{1}$. The permuted inverse quantum Fourier Transform subroutine \cite{Nielsen} transforms this state into $\ket{b_1}\otimes\ket{b_2}\otimes\cdots\otimes\ket{b_R}$. Thus, readout of the phase register retrieves a rational approximation to $\phi$.

\subsection{Quantum dispersion of the PEA readouts \label{subsec:PEA}}

Even if we have an ideal fault-tolerant quantum computer, there are two sources of uncertainty in the implementation of the PEA. First, if the eigenvalue $\phi$ is not exactly equal to a binary fraction representation $0.\phi$ then different such outputs can be obtained, with probabilities given by \cite{Nielsen}
\begin{align}
P_{\phi}(0.\phi) & =\begin{cases}
1 &,\;\phi=0.\phi\\
\frac{1}{2^{2R}}\frac{\sin^{2}\left(2^{R}\pi\left(\phi-0.\phi\right)\right)}{\sin^{2}\left(\pi\left(\phi-0.\phi\right)\right)} &,\;\phi\neq0.\phi
\end{cases},
\label{eq:PMF_PEA}
\end{align}
where $R$ is the number of resolution bits used. It is apparent that, as $R$ increases, the accuracy of the procedure improves.

The second caveat in the implementation of the PEA comes from the fact that the initial state on the simulation register might not be an eigenstate of ${\cal U}$, but a linear superposition 
\begin{equation}
\ket{\psi}=\sum_n c_n \ket{\phi_n}.
\label{ulin}
\end{equation}
In this case, the algorithm will collapse the wave function over one of the eigenstates, with a probability $|c_n|^2$, and provide the phase estimate corresponding to the eigenstate $\phi_n$ (see Appendix \ref{sec:QPE-super} for a demonstration).

In general, taking into account both sources of uncertainty, we can show (see Appendix \ref{sec:QPE-super}) that the probability of reading out a given phase $0.\phi$, for the state in Eq.~(\ref{ulin}), is given by 
\begin{equation}
P(0.\phi)= \sum_n |c_n|^2 P_{\phi_n}(0.\phi).
\label{eq:PMF_gen}
\end{equation}

These two sources of uncertainty make it necessary to run the PEA several times and analyze the relative frequency of the $0.\phi$ readouts, $P_\phi(0.\phi)$. In this work, we only consider the case of preparing $\ket{\psi}$ as an eigenstate of the ${\cal H}$ operator. This is the building block for understanding the simulation of superposition states.

\subsection{Circular statistics}

Consider the execution of the PEA for a given Hamiltonian, ${\cal H}$, using a fixed initial eigenstate $\ket{\phi}$, so that the eigenvalue of ${\cal U}$ is given by $e^{i 2\pi \phi}$. The first thing to note from Eq.~(\ref{eq:PMF_PEA}) is that the final statevector on the phase register subspace perfectly encodes $\phi$, and it can in principle be estimated from a sample with adequate confidence intervals, even with $R=1$. However, this may not be a feasible approach in the presence of noise, when the distribution is unknown, making it necessary to explore different estimators.

Let us take two different ways of inferring $\phi$ from this parent distribution. The first one would be to identify $\phi$ with the $0.\phi$ that has the largest $P_\phi(0.\phi)$, that is $\hat{\phi}=\widetilde{0.\phi}$ for $P_\phi(\widetilde{0.\phi})\geq4/\pi^{2}$. This biased estimator is the {\em majority} rule most often used in the literature \cite{cleve98, Nielsen, aspuru2005, omalley2016}, for which the absolute accuracy error is bounded by $\epsilon<2^{-(R+1)}$, where
\begin{equation}
\epsilon=| \hat{\phi}-\phi|.
\label{eq:acc}
\end{equation}

The alternative way is to take an average. For that matter, we define the {\em first trigonometric moment about the mean direction} \cite{MardiaJupp1999} as
\begin{equation}
\theta_{1} = \sum_{0.\phi}P_{\phi}(0.\phi)e^{i2\pi0.\phi} \equiv\rho e^{i2\pi\mu},
\label{eq:mrv}
\end{equation}
which using Eq.~(\ref{eq:PMF_PEA}) computes to
\begin{equation}
\theta_{1} =\frac{2^{R}-1}{2^{R}}e^{i2\pi\phi}+\frac{1}{2^{R}}e^{-i\left(2^{R}-1\right)2\pi\phi}.
\label{eq:mrv_PMF}
\end{equation}
Here, $\mu$ is the {\em mean phase direction}, given by
\begin{equation}
\mu\left(\phi\right)=\frac{1}{2\pi}\arctan\left(\frac{A\sin\left(2\pi\phi\right)-\sin\left(A2\pi\phi\right)}{A\cos\left(2\pi\phi\right)+\cos\left(A2\pi\phi\right)}\right),
\label{eq:mpd}
\end{equation}
with $A\equiv2^{R}-1$, properly piecewise defined such that $\mu\left(\phi\right) \in \left[0,1\right($ is continuous. The modulus of $\theta_1$,
\begin{equation}
\rho=\sqrt{4^{-R} \left(4^R - 2^{R+1} + 2 + 2 A \cos \left(2^{R+1} \pi \phi\right)\right)},
\label{eq:mrl}
\end{equation}
is called the {\em mean resultant length} and provides a measure of how narrow this unimodal distribution is. It can be translated to a more familiar form to non-circular statistics by defining the \emph{phase circular standard deviation} as
\begin{equation}
\sigma=\frac{\sqrt{-2\ln\rho}}{2\pi}.
\label{eq:PCSTD}
\end{equation}
We can assign a geometric interpretation to Eq.~(\ref{eq:mrv}): if we think of each complex number in the sum as a vector in the complex plane, $\rho$ stands for the length of the resulting vector, and $\mu$ describes its orientation. A uniform distribution would yield a vanishing $\rho$.

\begin{figure}
\includegraphics[width=1.\linewidth]{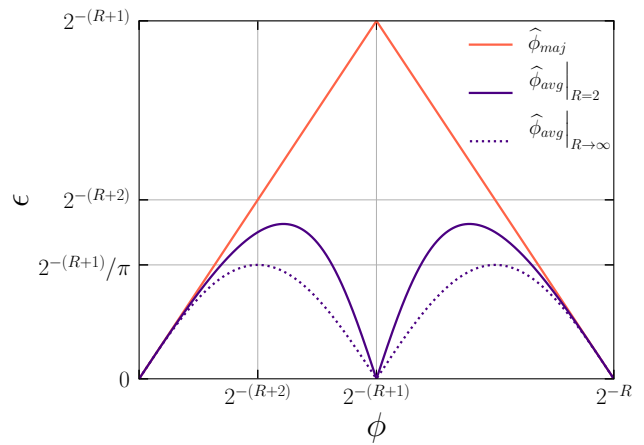}
\caption{Modulus of the accuracy error, defined in Eq.~(\ref{eq:acc}), as a function of the phase $\phi$ for two different phase estimators $\hat{\phi}$: the maximum of Eq.~(\ref{eq:PMF_PEA}) (orange) and the mean phase direction of the distribution obtained with Eq.~(\ref{eq:mpd}) (purple). $R$ stands for the number of register qubits. All the curves for $\hat{\phi}_{avg}$ with $R>2$ lie between the filled and the dashed purple lines.}
\label{fig:acc}
\end{figure}

From this approach, we can define a new estimator $\hat{\phi}=\mu$. The maximum accuracy error can be obtained straightforwardly and is seen to be bounded by $\epsilon<2^{-(R+2)}$ for $R\geq2$, decreasing monotonically with $R$. Therefore, this is again a biased estimator for general $\phi$ but significantly improves on the previous rule. Specifically, the estimation of $\phi$ from the mean phase direction using $R$ qubits is below the accuracy error bound for the majority rule estimator with $R+1$ qubits. Therefore, in this sense, the estimation of the phase using the average permits to decrease $R$ by one unit, which makes circuits shallower by avoiding one exponentiation of ${\cal U}$.

In Fig.~\ref{fig:acc}, we compare the absolute value of the bias for the majority and average rules as we ramp $\phi$ between two adjacent values of $0.\phi$. It is apparent that $\epsilon$ is always {\em smaller} for the average rule. This is in part due to a cancellation of errors: when the phase $\phi$ is right in the middle, the average rule gives an exact estimation. Interestingly, an asymptotic limit is hit for infinite resolution, $R\rightarrow\infty$, in which the error bound scales as $2^{-(R+1)}/\pi$.

We also note that Eq.~(\ref{eq:mpd}) reveals there is a one-to-one map between $\phi$ and $\mu$, such that Eq.~(\ref{eq:PMF_PEA}) can be re-parametrized by the mean phase direction. Given that $\mu$ is simple to estimate from the sample, inverting Eq.~(\ref{eq:mpd}) provides an unbiased estimator for $\phi$ at any $R\geq2$. This allows sparing increasing executions of the unitary and trade them by sampling, which represents a meaningful improvement for the execution of QPE algorithms on devices with short coherence time, such as NISQ hardware.

Using any of these estimators for $\phi$ would be useful if we were only interested in determining the phase of the propagator, however, we want to get to the eigenvalues of the Hamiltonian. For that, we perform several experiments with different $\tau$ to employ Eq.~(\ref{eq:eig_est}), and in this case, we can just as well use $\mu$ directly without the additional inversion step.

\subsection{Sampling \label{sec:sampling}}

Running the PEA in a fault-tolerant quantum computer requires taking a finite sample of $O$ observations from the parent distribution in Eq.~(\ref{eq:PMF_PEA}) to infer its single parameter: either $\phi$ or $\mu$ if the distribution is re-parametrized. We can use the sample mean phase direction, $\overline{\mu}$, as an unbiased estimator of $\mu$ given by Eq.~(\ref{eq:mpd}). If the same experiment could be repeated with $S$ samples of $O$ observations each to compute $\overline{\mu}$ in each one, we would approximate the sampling distribution of $\hat{\mu}$ for the chosen value of $O$. This could in turn be used to calculate $\sigma_{\hat{\mu}}$, given by the circular standard deviation of the $S$ sample mean phase directions $\overline{\mu}$.

\begin{figure}
\includegraphics[width=1.\linewidth]{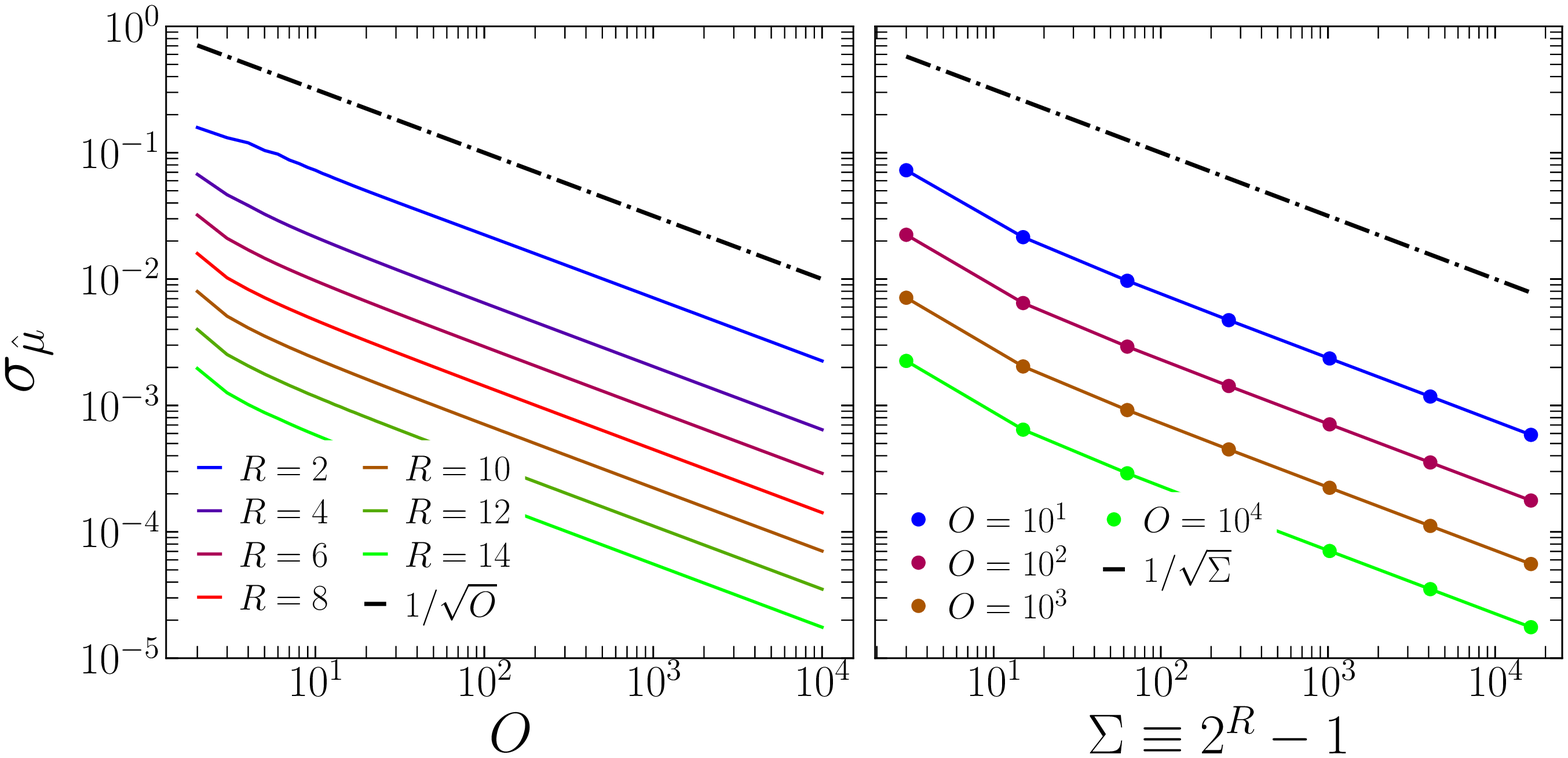}
\caption{Numerical predictions for $\sigma_{\hat{\mu}}$ obtained with $S=10^7$ samples drawn from Eq.~(\ref{eq:PMF_PEA}) at $\phi=2^{-(R+1)}$, where dispersion is maximal. Most other $\phi$ values yield $\sigma_{\hat{\mu}}$ lower than these bounds. On the left panel, the dependence of $\sigma_{\hat{\mu}}$ on $O$ is traced for different values of $R$. The right panel shows how increasing $R$ reduces $\sigma_{\hat{\mu}}$ while fixing the number of observations. $\Sigma$ represents the total number of $\cal{U}$ executions throughout the PEA. Recall that $\mu$ is different for each $R$.}
\label{fig:pcsd_mpd}
\end{figure}

In the left and right panels of Fig.~\ref{fig:pcsd_mpd}, we verify scaling of $\sigma_{\hat{\mu}}$ with the inverse square root of $O$ and $\Sigma$, respectively. For the case of the right panel, $\sigma_{\hat{\mu}}$ can actually decrease faster when the exponentiation of the unitary is implementable with fewer executions. As we show below for the Zeeman and Ising Hamiltonians, the exponentiation of these unitaries can be implemented with a single parametrized execution using the IBM~Q gate set.

Quantifying the standard error of $\hat{\mu}$ in this way requires a large quantum computational overhead. Fortunately, there are less computationally intensive ways of doing so, for instance through bootstrapping \cite{EfronHastie2016}. This approach can be used straightforwardly with a fault-tolerant quantum computer to quantify the dispersion of $\hat{\mu}=\overline{\mu}$ and, consequently, its {\em mean squared error}
\begin{equation}
\mathrm{MSE}\left(\hat{\mu}\right)=\sigma_{\hat{\mu}}^{2}+\mathrm{bias}^{2}_{\hat{\mu}},
\label{eq:MSE-MPD}
\end{equation}
since the bias
\begin{equation}
\mathrm{bias}_{\hat{\mu}}=\left\langle \hat{\mu}\right\rangle-\mu
\label{eq:bias-MPD}
\end{equation}
would be zero in these computers.

However, in state-of-the-art machines, the experimental distribution of results does not approach the unitary one because of the introduction of errors which can generally add both dispersion and skewness to Eq.~(\ref{eq:PMF_PEA}). Since bootstrapping only requires the sequence of observations to be independent and identically distributed, we can still quantify dispersion of the estimator with this technique if the noise process is stationary. This is useful for making no assumptions on a noise model besides this one. Even so, dispersion is no longer enough to account for the $\mathrm{MSE}\left(\hat{\mu}\right)$ because the noise fluctuations can bias $\hat{\mu}$, frequently by a much larger magnitude than $\sigma_{\hat{\mu}}$.

On that account, for the results obtained in state-of-the-art devices, we consider the $0.\phi$ phase circular standard deviation of the sample as the error-bar for the $\hat{\mu}$ estimation. This choice is an overestimation of the experimental $\mathrm{MSE}\left(\hat{\mu}\right)$, but it is a safe strategy. As we explain below, we can improve accuracy and precision of the final eigenvalue estimation by performing several experiments with different values of the control parameter $\tau$ in the propagator and fitting the results to the expected theoretical model.

\subsection{Iterative PEA}

Quantum phase estimation based on the $R$-qubit quantum Fourier transform can lead to a very high number of gates due to the need to implement the $R$ powers of $\mathcal{U}$ in a single long-depth circuit. To avoid this, we can separate the simulation of each of these powers into different circuits and perform QPE with only an ancillary qubit and a $R$-fold iterative procedure. This measurement-based approach is made possible by the semiclassical quantum Fourier transform \cite{griffiths96} and is known as the iterative phase estimation algorithm (IPEA).

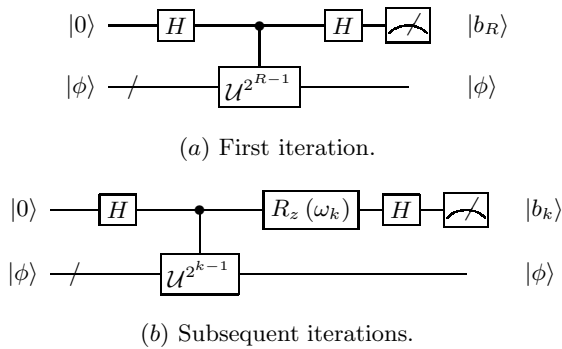
\begin{figure}
\begin{centering}
\begin{tabular}{c}
\centering{\
\Qcircuit @C=1em @R=1em {
   \lstick{\ket{0}} & \qw & \gate{H} & \ctrl{1} & \gate{H} & \meter & \rstick{\ket{b_R}} \\
   \lstick{\ket{\phi}} & \qw {/} & \qw & \gate{\mathcal{U}^{2^{R-1}}} & \qw & \qw & \rstick{\ket{\phi}}} \par}\tabularnewline
\tabularnewline
$\left(a\right)$ First iteration.\tabularnewline
\tabularnewline
\centering{\
\Qcircuit @C=1em @R=1em {
   \lstick{\ket{0}} & \qw & \gate{H} & \ctrl{1} & \gate{R_{z}\left(\omega_{k}\right)} & \gate{H} & \meter & \rstick{\ket{b_k}} \\
   \lstick{\ket{\phi}} & \qw {/} & \qw & \gate{\mathcal{U}^{2^{k-1}}} & \qw & \qw & \qw & \rstick{\ket{\phi}}} \par}\tabularnewline
\tabularnewline
$\left(b\right)$ Subsequent iterations.\tabularnewline
\tabularnewline
\end{tabular}
\par\end{centering}
\caption{Quantum circuits used by the iterative phase estimation algorithm.}
\label{fig:IPEA}
\end{figure}

In the IPEA, each digit of one $0.\phi$ observation is estimated with a dedicated measurement using the circuits in Fig.~\ref{fig:IPEA}. For that, we first fix a desired resolution $R$ in Eq.~(\ref{eq:binrepr}) and adopt one of two exploration procedures for choosing between the intermediate measured states produced by these circuits: {\em exhaustive} or {\em non-exhaustive}, explained below.

To begin with, we perform a given number of executions of the circuit in Fig.~\ref{fig:IPEA}(a) to obtain the relative measurement frequencies of the two basis states of the ancillary qubit. These are used to decide on the $b_R$ bit, in accordance with the adopted exploration procedure. Having fixed this digit, we start an iterator variable $k$ decreasing from $R-1$ to $1$. For each $k$, the circuit in Fig.~\ref{fig:IPEA}(b) is repeatedly executed with
\begin{equation}
\omega_{k}=-2\pi\sum_{l=2}^{R-k+1}\frac{b_{k+l-1}}{2^{l}}
\label{eq:IPEA-angle}
\end{equation}
and the ancillary qubit is measured to fix $b_k$, again from the obtained histogram. In this way, we determine all the bits of Eq.~(\ref{eq:binrepr}) from the least to the most significant, at each iteration performing the $z$-rotation with an angle that is a function of all the previously measured bits.

The two different modes of operating the IPEA differ in the exploration policy of the state space of the binary tree encoding all the $0.\phi$ decompositions of the phase. With {\em exhaustive} IPEA, we can probe the probability mass function from Eq.~(\ref{eq:PMF_PEA}) over all the $2^R$ binary strings, just like with the PEA. Therefore, this exploration mode allows us to also employ Eq.~(\ref{eq:mpd}) when initializing the algorithm with $\mathcal{H}$ eigenstates, as well as to explore general states.

\begin{figure}
\includegraphics[width=1.\linewidth]{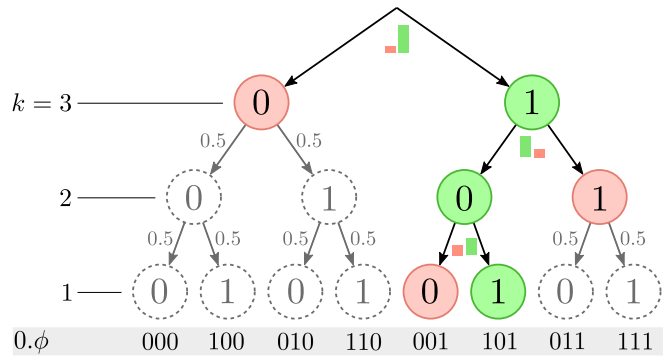}
\caption{Reconstruction procedure of the LPMF over all binary strings performed with non-exhaustive IPEA, illustrated for the case of $R=3$. Only one branch of the full binary tree is explored in this algorithm. Green nodes are the digits obtained from a majority rule at each iteration, which encode the final estimation. Branches below the red nodes are unexplored.}
\label{fig:non-exhaustive-IPEA}
\end{figure}

Sometimes, we are only interested in obtaining the value $\widetilde{0.\phi}$ for which the probability of  Eq.~(\ref{eq:PMF_PEA}) is maximal. For that, we can perform a {\em non-exhaustive} exploration of the state space using the IPEA as implemented by Dob\v{s}\'{\i}\v{c}ek et al. \cite{Dobsicek07}. This procedure consists in running each iteration a sufficient number of times to let probabilities converge, choose the most frequently obtained bit as $b_k$ and proceed to the next iteration. The final estimation $\hat{\phi}=\widetilde{0.\phi}$ is the binary string immediately constructed with the most frequent bits obtained in all the iterations. This procedure is equivalent to the PEA with the majority rule. Since this only explores one branch of the full binary tree of the $R$-bit string (see Fig.~\ref{fig:non-exhaustive-IPEA}), it cannot be used to obtain complete information of input superpositions of $\mathcal{U}$ eigenstates.

We again emphasize that our discussion of the mean phase direction estimator is directly applicable to the exhaustive IPEA. However, in the experiments here reported, we only use the non-exhaustive version of the IPEA for the reasons explained in section \ref{sec:hubres}. In this scenario, we can take into account all the 2-state histograms at the intermediate steps to construct the sample approximation to a lossy version of the full probability mass function given by Eq.~(\ref{eq:PMF_PEA}) over all the possible $R$-bit strings. To approximate this lossy probability mass function (LPMF), we organize the relative frequencies of measurement obtained for each bit in a binary tree format where a single full-depth branch encodes $\widetilde{0.\phi}$ and the unexplored bit measurements correspond to chopped nodes at every level. 

Then, the relative frequencies of the unexplored nodes are uniformly propagated down to their descendant leaves in the tree to reconstruct the sample approximation to the LPMF, which in turn approximates Eq.~(\ref{eq:PMF_PEA}). This LPMF distribution cannot provide sufficient information to discern the complexity of Eq.~(\ref{eq:PMF_PEA}) across all the different $R$-bit states, but allows computing the phase circular standard deviation with Eq.~(\ref{eq:PCSTD}), used as a measure of dispersion of the sample obtained with non-exhaustive IPEA. Using Eq.~(\ref{eq:mpd}) over the LPMF cannot improve accuracy.

The IPEA has the big advantage of achieving a very large reduction of the gate count while also overcoming the limit in resolution imposed by the number of qubits used to register the phase in the PEA. The only significant increase in computing time comes from preparing the initial state at every run.

\subsection{Least squares regression}

In order to determine a given eigenvalue, we use Eq.~(\ref{eq:eig_est}). To do so, we carry out simulations with different values for $\tau$ and extract the slope by fitting the phase estimates $\hat{\phi}$ to a theoretical model $f(\tau)$. In the case of the majority rule estimator, we take $f(\tau) = m\tau + b$ because even if this is not the exact model, there is a linear dependence of $\hat{\phi}$ in $\tau$. In the case of the average rule estimator, we use $f(\tau)=\mu(m\tau + b)$, where $\mu$ is given by Eq.~(\ref{eq:mpd}). However, given the periodic nature of the data, which causes problems when $\hat{\phi}$ is close to 0 and 1, our approach is to use the least squares fitting method to minimize a modified $\chi^2$ functional given by
\begin{equation}
\chi^2_\text{circ} = \sum_{i=1}^{N} 
\frac{|e^{i 2 \pi \hat{\phi}_i}-e^{i 2 \pi f(\tau_i)}|^2}{\sigma_i^2},
\label{eq:X2circ}
\end{equation}
where $\sigma_i$ is the error-bar associated to the estimate for $\tau_i$. The fitting procedure determines the parameters $m$ and $b$. This functional can be split into two,
\begin{equation}
\chi^2_\text{circ} = \chi^2_{\cos} + \chi^2_{\sin},
\end{equation}
where
\begin{equation}
\chi^2_{\cos} = \sum_{i=1}^{N} 
\left(
\frac{\cos( 2 \pi \hat{\phi}_i)-\cos(2 \pi f(\tau_i))}
{\sigma_i}
\right)^2,
\label{eq:X2cos}
\end{equation}
\begin{equation}
\chi^2_{\sin} = \sum_{i=1}^{N} 
\left(
\frac{\sin( 2 \pi \hat{\phi}_i)-\sin(2 \pi f(\tau_i))}
{\sigma_i}
\right)^2.
\label{eq:X2sin}
\end{equation}
Thus, minimizing the functional in Eq.~(\ref{eq:X2circ}) is equivalent to using the standard least squares method to minimize $\chi^2_{\cos}$ and $\chi^2_{\sin}$ simultaneously.

\section{Spin Hamiltonians\label{HAM}}

In order to test our DQS and QPE methods, we first apply them to spin-$1/2$ Hamiltonians. For such problems, the eigenstates of the Hamiltonian are identical to the  computational  basis. Here, we introduce the two spin models considered in this work and show explicitly how to implement the unitary evolution operators using one- and two-qubit gates.

\subsection{Two-level system}
The simplest Hamiltonian one could think of is a two-level system with energy splitting $\omega$,
\begin{equation}
{\cal H}_{\rm Z}= \omega Z,
\label{TLS}
\end{equation}
where $Z$ is the Pauli $Z$ operator. The Hilbert space has dimension 2 and can therefore be encoded in a single qubit. In  the spin language, this model can be interpreted as the Zeeman Hamiltonian of a $S=1/2$ spin. The energy levels of ${\cal H}_{\rm Z}$ are $\pm \omega$.

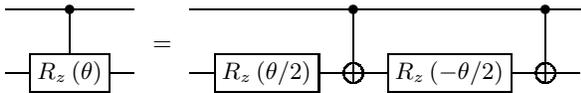
\begin{figure}[h!]
\noindent \begin{centering}
{\footnotesize{}}
\begin{tabular}{ccc}
\centering{\
\Qcircuit @C=1em @R=1.7em {
   & \ctrl{1} & \qw \\
   & \gate{R_{z}\left(\theta\right)} & \qw } \par} & $\begin{array}{c}
\\
\\
\\
=
\end{array}$ & \Qcircuit @C=1em @R=1.7em {
   & \qw & \ctrl{1} & \qw & \ctrl{1} & \qw \\
   & \gate{R_{z}\left(\theta/2\right)} & \targ  & \gate{R_{z}\left(-\theta/2\right)} & \targ & \qw }\tabularnewline
\end{tabular}{\footnotesize\par}
\par\end{centering}
\caption{Gate decomposition of the controlled-$R_{z}\left(\theta\right)$ operator. This circuit directly implements the controlled propagator for the two-level system.
}
\label{fig:cRz}
\end{figure}

For the two-level system, we take the qubit computational basis as the eigenstate basis. Therefore, the {\em initialization} is straightforward. The unitary evolution operator can be simulated in the quantum computer using the phase shift gate as
\begin{equation}
{\cal U}_{\rm Z}(\tau)=e^{- i{\cal H}_{\rm Z} \tau}
= R_z(\theta)=
\left(\begin{array}{cc} e^{-i\theta/2} & 0 \\ 
0 & e^{i\theta/2}\end{array}\right),
\end{equation}
where $\theta = 2\omega \tau$. However, to control the action of ${\cal U}_{\rm Z}(\tau)$, we need to implement two-qubit gates. In Fig.~\ref{fig:cRz}, we show how to decompose a controlled-$R_z(\theta)$ operation in terms of primitive gates.

\subsection{Ising dimer}

Our second model moves one step up in complexity scale and accounts for 2 spins with Ising-like interactions that preserve the spin operator $S_z$. The Hamiltonian reads as 
\begin{equation}
{\cal H}_{\rm I} = \omega_{1} Z_{1} + \omega_{2} Z_{2} +  \omega_{J} Z_{1} Z_{2},
\label{Isingdimer}
\end{equation}
where the first two terms correspond to decoupled Zeeman Hamiltonians for the two spins, whereas the third term accounts for a Ising-like exchange interaction with energy $\omega_J$.

The four eigenvalues of this Hamiltonian are $\varepsilon_{s_1,s_2}=s_1\omega_1 + s_2\omega_2+s_1s_2\omega_J$, where $s_{1,2}=\pm1$. As in the case of the two-level system, the eigenstates of ${\cal H}_{\rm I}$ can also be encoded in the computational basis of the quantum computer. This is done with a trivial mapping between the state with eigenvalue $\varepsilon_{s_1,s_2}$ and the qubit state $\ket{\frac{1-s_{1}}{2},\frac{1-s_{2}}{2}}$.

\begin{figure}[h]
\centering{\
\Qcircuit @C=1em @R=1.2em {
   \lstick{} & \ctrl{2} & \ctrl{3} & \qw & \ctrl{3} & \qw & \qw \\
   & & & & & \\
   \lstick{q_{1}:} & \gate{R_z(\theta_{1})} & \qw &  \ctrl{1} & \qw & \ctrl{1} & 
\qw \\
  \lstick{q_{2}:} & \qw & \gate{R_z(\theta_{2})} & \targ & \gate{R_z(\theta_{3})} & \targ & \qw 
\gategroup{3}{4}{4}{6}{1em}{--}
} \par}
\caption{Quantum circuit to implement the controlled unitary evolution operator for the Ising model. The gate decomposition of the controlled-$R_z(\theta)$ gates is shown in Fig.~\ref{fig:cRz}.} 
\label{fig:cU-IS}
\end{figure}
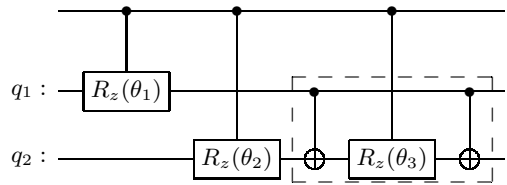

The Hamiltonian for the Ising dimer, Eq.~(\ref{Isingdimer}), is a sum of mutually commuting terms, so that its unitary evolution operator can be decomposed as a product of operators corresponding to each term. In the quantum computation language, it can be written as
\begin{equation}
{\cal U}_{\rm I}(\tau)=e^{-iH_{\rm I}\tau}= R_z^{(1)}(\theta_1) R_z^{(2)}(\theta_2) e^{-i\theta_3 Z_1 Z_2/2},
\end{equation}
where $\theta_1=2\omega_1 \tau$, $\theta_2=2\omega_2 \tau$ and $\theta_3=2\omega_J \tau$. The quantum circuit that implements the controlled-${\cal U}_{\rm I}(\tau)$ operation is shown in Fig.~\ref{fig:cU-IS}, where the implementation of the term $e^{-i\theta_3 Z_1 Z_2/2}$ is boxed.

\section{Results for the spin Hamiltonians \label{sec:spinres}}

We now present the experimental results for the spin models described in the previous section. All digital quantum simulations were carried out on the IBM Q hardware. In this platform, single-qubit operations are implemented with just three physically parametrized gates,
\begin{equation}
\begin{split}U_{1}(\lambda) & =\begin{pmatrix}1 & 0\\
0 & e^{i\lambda}
\end{pmatrix},\\
U_{2}(\phi,\lambda) & =\frac{1}{\sqrt{2}}\begin{pmatrix}1 & -e^{i\lambda}\\
e^{i\phi} & e^{i(\lambda+\phi)}
\end{pmatrix},\\
U_{3}(\theta,\phi,\lambda) & =\begin{pmatrix}\cos\left(\theta/2\right) & -e^{i\lambda}\sin\left(\theta/2\right)\\
e^{i\phi}\sin\left(\theta/2\right) & e^{i(\lambda+\phi)}\cos\left(\theta/2\right)
\end{pmatrix},
\end{split}
\label{eq:ibmgateset}
\end{equation}
which together with the c$X$ gate, form a universal gate set only limited by noise and decoherence. In particular, our implementations use the \emph{ibmq\_20\_tokyo} 20-qubit device. Even though this machine offers 20 qubits, they are not fully connected, meaning that we cannot make direct c$X$ operations between any two arbitrary qubits. This is an important issue in NISQ computing, as it can greatly increase the gate count. For that reason, for both the two-level system and the Ising dimer, we took a subset of 4 fully-connected qubits that \emph{ibmq\_20\_tokyo} offers. This allowed us to end up with quantum circuits with small enough depth, for which the PEA can still be employed. Naturally, this choice imposes a limited resolution. For the two-level system, we only need 1 qubit to simulate time evolution and we are left with $R=3$ qubits in the phase register. For the Ising dimer, the simulation register requires 2 qubits and therefore we can only use $R=2$ qubits to register the phase. Given this relatively small resolution in the PEA, we compare the experimental results with those from the ideal distribution expected as in Eq.~(\ref{eq:PMF_PEA}).

\begin{figure}
\includegraphics[width=1.\linewidth]{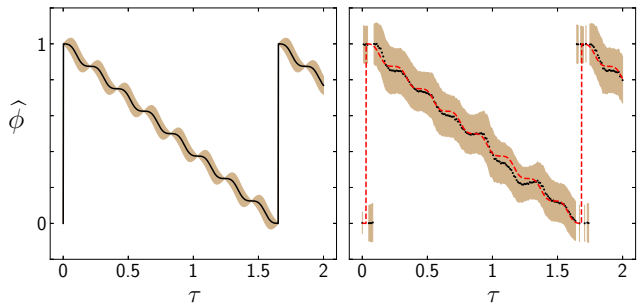}
\caption{Results for the PEA implementation of the two-level system. On the left-side panel, the black line traces the mean phase direction parameter of the parent distribution in Eq.~(\ref{eq:PMF_PEA}) as given by Eq.~(\ref{eq:mpd}) with $\phi=-\varepsilon\tau/(2\pi)$; the brown region enclosing $\mu$ corresponds to $1\sigma$, computed with Eq.~(\ref{eq:PCSTD}). On the right-side panel, the experimental results obtained in \emph{ibmq\_20\_tokyo} are shown. The black dots plot $\overline{\mu}$ while the brown region enclosing each point is one experimental $\sigma$. The dashed red line is given by $f(\tau) = \mu (m\tau+b \pmod{1})$, where $m=-0.6041\pm0.0039~$ and $b=0.0174\pm0.0046$ are fitting parameters. The corresponding $\chi^2$ functional per number of degrees of freedom is $\chi_{\text{circ}}^2/\text{ndf} = 0.16$.}
\label{fig-Z}
\end{figure}

In Fig.~\ref{fig-Z}, we show the results for the mean phase direction and the phase circular standard deviation obtained for the two-level system, fixing $\omega = 3.8$ and the initial state as $\ket{\psi}=\ket{0}$, whose eigenvalue is $\varepsilon = \omega$. The total number of gates in this circuit is 29, of which 12 are c$X$. We perform 200 experiments for different values of $\tau\in\left[0,2\right]$, in each one taking $O=8192$ shots. We observe that, even with just $R=3$, the quantum dispersion of the PEA is much smaller than the experimental noise.

For the reasons explained in section \ref{sec:sampling}, we fit the experimental $\overline{\mu}\left(\tau\right)$ with $\sigma$ given by the sample phase circular standard deviation in Eq.~(\ref{eq:X2cos}) and Eq.~(\ref{eq:X2sin}). We find a good agreement between experimental and theoretical results, obtaining an estimation for the eigenvalue of $\hat{\varepsilon}=3.80\pm 0.02$. The fact that $b=0.0174\pm0.0046$ does not cover the expected value $b=0 \bmod{1}$ is interpreted as a phase shift error that can be due to calibration, for instance.

\begin{figure}
\includegraphics[width=1\linewidth]{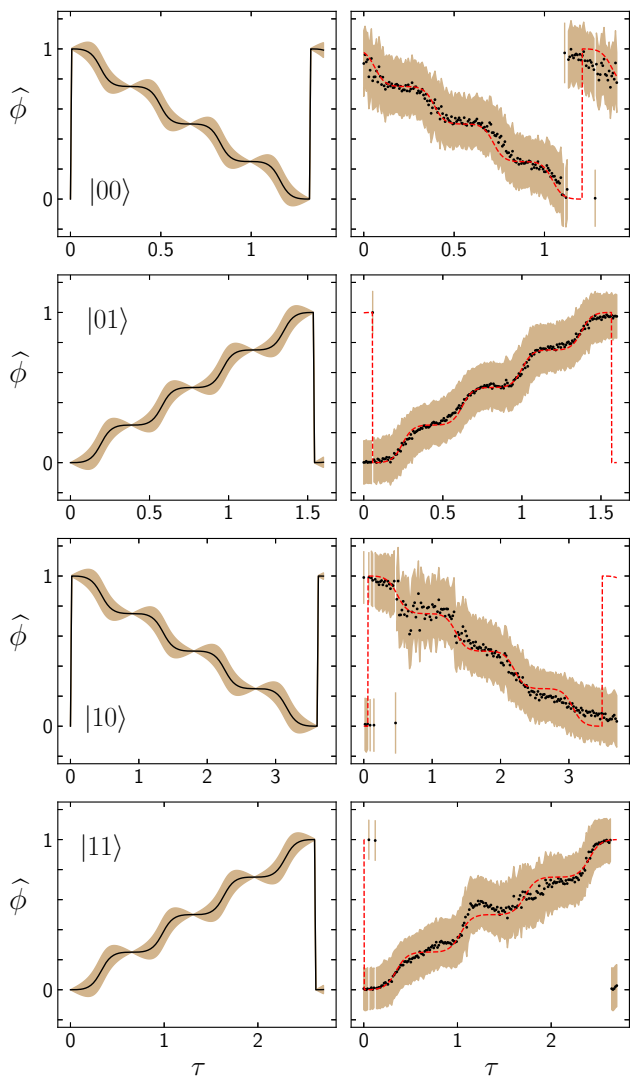}
\caption{PEA results for the simulation of the four eigenstates of the Ising dimer. The left-side column plots $\mu$ (black) and $1\sigma$ (brown) derived from the parent distributions of fault-tolerant circuits, as in the case of the Zeeman simulation. The right-side panels show \emph{ibmq\_20\_tokyo} results: $\overline{\mu}$ is represented by the black dots and the experimental disperson of $1\sigma$ is marked in brown; dashed red lines show the fit results of $f(\tau) = \mu (m\tau+b \pmod{1})$ to this data, with $m$ and $b$ as parameters (see Table~\ref{tab:IS}).}
\label{fig-IS}
\end{figure}

In the case of the Ising dimer, we explore the initialization of the system in its four eigenstates, $\ket{00}$, $\ket{01}$, $\ket{10}$ and $\ket{11}$. These are explored in separate, though, and we do not address here the linear superposition case. For computation purposes, we take $\omega_1 = 0.33$, $\omega_2 = 3.24$ and $\omega_J = 1.17$. The results, obtained with $O=5000$ shots, are shown in Fig.~\ref{fig-IS} and Table~\ref{tab:IS}. Despite the different initialization of each eigenstate, which in this case takes at most two $X$ gates for the preparation of $\ket{11}$, all the PEA circuits for the Ising dimer are compiled by the platform to yield 35 gates, of which 18 are c$X$. Once again, we observe that the dispersion due to finite $R$ is much smaller than the experimental noise. For this model Hamiltonian, the increased complexity leads to experimental values that do not cover the theoretical ones. Even so, the relative errors are small and the standard score is roughly between 1-3 (except for the $\ket{11}$ eigenstate), meaning that the experimental values fall off from the theoretical ones by 1-3 error bars.
Therefore, we verify that the PEA can handle the Ising dimer. Finally, we note that for three out of four initial states, we get values of $b$ that indicate the presence of phase shift errors. In particular, when the state is initialized as $\ket{00}$, this phase shift is evident just by comparing the top plots of Fig.~\ref{fig-IS}. The phase shift error, though, does not affect the estimation of the energies.

\begin{table}
\newcolumntype{R}{>{\centering\arraybackslash}X}
\noindent
\begin{tabularx}{\columnwidth}{ *{5}{R} }
\hline
\hline
 $\ket{u}$ & $\left|00\right\rangle $ & $\left|01\right\rangle $ & $\left|10\right\rangle $ & $\left|11\right\rangle$ \smallskip{}\\
\hline
$\varepsilon$ & $4.74$\smallskip{}
 & $-4.08$ & $1.74$ & $-2.40$\\
\multirow{2}{*}{$m$} & $-0.763$ & $0.662$ & $-0.292$ & $0.364$\\
 & $\pm0.009$\smallskip{}
 & $\pm0.007$ & $\pm0.004$ & $\pm0.003$\\
\multirow{2}{*}{$b$} & $-0.078$ & $-0.037$ & $0.018$ & $-0.001$\\
 & $\pm0.007$\smallskip{}
 & $\pm0.007$ & $\pm0.007$ & $\pm0.005$\\
$\chi_{\text{circ}}^2/\text{ndf}$ & $0.37$\smallskip{}
 & $0.09$ & $0.38$ & $0.28$\\
\multirow{2}{*}{$\hat{\varepsilon}$} & $4.79$ & $-4.16$ & $1.83$ & $-2.29$\\
 & $\pm0.06$\smallskip{}
 & $\pm0.05$ & $\pm0.03$ & $\pm0.02$\\
$\left(\varepsilon-\hat{\varepsilon}\right)/\left|\varepsilon\right|$ & $-0.01$\smallskip{}
 & $0.02$ & $-0.05$ & $-0.05$\\
$\left(\varepsilon-\hat{\varepsilon}\right)/\delta\hat{\varepsilon}$ & $-0.8$ & $1.6$ & $-3.0$ & $-5.5$\\
\hline
\hline
\end{tabularx}
\caption{Fitting parameters for the experimental results of the PEA with the Ising dimer.}
\label{tab:IS}
\end{table}

\section{Hubbard model\label{sec:hub}}

We now consider a 2-site Hubbard model, whose Hamiltonian reads as
\begin{equation}
\mathcal{H}_{\rm H}= U\sum_{i=a,b} n_{i\uparrow}n_{i\downarrow} - t \sum_{\sigma=\uparrow,\downarrow} \left(a^{\dagger}_{\sigma}b_{\sigma}+b^{\dagger}_{\sigma} a_{\sigma} \right), 
\end{equation}
where $U, t$ are positive parameters, $a_\sigma/b_\sigma$ is the fermionic annihilation operator for an electron in site $a/b$ with spin $\sigma=\uparrow,\downarrow$ and $n_{a\sigma} = a^{\dag}_{\sigma}a_{\sigma}, n_{b\sigma} = b^{\dag}_{\sigma}b_{\sigma}$ are the corresponding number operators.

The DQS of the Hubbard model poses several additional challenges missing in the case of the two spin models. First, we need to map fermions to qubits. Second, the model is a sum of non-commuting terms, which complicates the simulation of the unitary evolution. Third, the preparation of the initial states is no longer trivial.

\subsection{From fermions to qubits: Jordan-Wigner transformation}

We use a Jordan-Wigner transformation \citep{jordan1928} to map fermionic operators into qubit operators. This requires one qubit per spin-site, which leads to 4 qubits in the case of the Hubbard dimer.
We associate the states $a\uparrow$, $b\uparrow$, $a\downarrow$, $b\downarrow$ (whose occupation can be either 0 or 1) to the qubits $1,2,3,4$, respectively. By doing so, the Hubbard Hamiltonian reads as $\mathcal{H}_{\rm H} = \mathcal{H}_U + \mathcal{H}_t$, with
\begin{equation}
H_{U} =\frac{U}{4}\left(2\boldsymbol{1} + Z_3 Z_1 + Z_4 Z_2 -\sum_{i=1}^4 Z_i\right),
\end{equation}
\begin{equation}
H_{t} =-\frac{t}{2}\left(X_2 X_1 + Y_2 Y_1 + X_4 X_3 + Y_4 Y_3 \right),
\end{equation}
where $X$ ($Y$) denotes the Pauli $X$ ($Y$) matrix, that acts on the qubit subspace labeled by the subscript.

The Hubbard dimer is thus mapped into a spin model with 11 terms that, unlike the case of the Ising dimer in Eq.~(\ref{Isingdimer}), are non-commuting. This brings an additional layer of complexity when it comes to simulate the unitary evolution. The quantum circuit to implement the PEA with this Hamiltonian, for just one $1^{\rm st}$-order Trotter step, with 3 register qubits and in a quantum computer with $3+4$ fully-connected qubits, already has close to 500 gates in total, of which over 200 are $\mathrm{c}X$ gates. Moreover, these numbers do not even account for the circuit that needs to be built in order to prepare the initial state. Both of these requirements make its practical implementation in current hardware impossible. Therefore, we adopt a different strategy, which requires a much smaller number of gates.

\subsection{Hubbard dimer at half-filling: compact representation}
At half-filling, the Hilbert space of the Hubbard dimer has 6 states. Two of them have spin projection $s_z=+1,-1$ and, since the spin operator $S_z$ commutes with the Hamiltonian, these are decoupled from the remaining four states, that have $s_z=0$. Thus, it is possible to find a compact representation \cite{moll16} for the non-trivial states of the half-filled Hubbard dimer that makes use of only 2 qubits. The four non-trivial states, written in the basis $\{a\uparrow$, $b\uparrow$, $a\downarrow$, $b\downarrow\}$, are mapped into qubit states as
\begin{align}
\{ 1,0,1,0 \} \equiv \ \uparrow\downarrow\bigcirc &= \ket{00} \nonumber \\
\{ 0,1,1,0 \} \equiv \ \ \downarrow  \ \ \uparrow \; &= \ket{01} \nonumber \\
\{ 1,0,0,1 \} \equiv \ \ \uparrow \ \ \downarrow \;  &= \ket{10} \nonumber \\
\{ 0,1,0,1 \} \equiv \bigcirc\uparrow\downarrow &= \ket{11}.
\label{reduced}
\end{align}
In this basis, the (reduced) Hamiltonian is given by
\begin{equation}
    \mathcal{H}'_{\rm H} = 
    \begin{bmatrix}
    U & -t & -t & 0 \\
    -t & 0 & 0 & -t \\
    -t & 0 & 0 & -t \\
    0 & -t & -t & U
    \end{bmatrix}.
    \label{eq:44}
\end{equation}

Now, we can encode this Hamiltonian using 2 qubits, $1$ and $2$, that act on the spin $\downarrow$ and $\uparrow$ electrons, respectively. Inspection of Eq.~(\ref{reduced}) shows that that $X_1$, for instance, carries out the hopping  the $\uparrow$ electron, whereas the operator $Z_1$ measures in which of the two Hubbard sites it is located.  Using these operators, we can write up the Hamiltonian in Eq.~(\ref{eq:44}) as
\begin{equation}
    {\cal H}'_{\rm H} = -t(X_1 +X_2) +\frac{U}{2}(Z_1 Z_2 + \boldsymbol{1})
    \label{eq:hub2}
\end{equation}
where $X=\left(\begin{array}{cc} 0&  1\\ 1 & 0\end{array}\right)$ is the Pauli $X$ operator. This permits to model the two-site Hubbard model at half-filling using only 2 qubits. After this reduction procedure, it must be noted that the Pauli operators of the Hamiltonian in Eq.~(\ref{eq:hub2}) are not directly related to the original fermions via a Jordan-Wigner transformation. It must also be noted that Eq.~(\ref{eq:hub2}) can represent a 2-site Ising model with transverse magnetic field.

The energy spectrum of Eq.~(\ref{eq:44}) is given by
\begin{eqnarray}
\varepsilon_{\pm} =\frac{U}{2}  \pm \sqrt{4t^2 +U^2/4}, \nonumber\\
\varepsilon_b=0\; \;\;  , \varepsilon_c=U
\label{eq:hub2-eig}
\end{eqnarray}
In the $t=0$ limit, these states have energies $\varepsilon_+=U=\varepsilon_c$, $\varepsilon_-=\varepsilon_b=0$. In the $U=0$ limit, we have $\varepsilon_b=\varepsilon_c=0$ and $\varepsilon_{\pm}=\pm 2 t$.
The corresponding eigenstates are
\begin{eqnarray}
|\varepsilon_b\rangle= \frac{1}{\sqrt{2}}\left(\ket{10}-\ket{01}\right),
\end{eqnarray}
\begin{eqnarray}
|\varepsilon_c\rangle= \frac{1}{\sqrt{2}}\left(\ket{11}-\ket{00}\right),
\end{eqnarray}
\begin{eqnarray}
\ket{\varepsilon_\pm}=
\alpha_{\pm}\left[\left|00\right\rangle +\left|11\right\rangle +\beta_{\pm}\left(\left|10\right\rangle +\left|01\right\rangle \right)\right],
\end{eqnarray}
where
\begin{equation}
\alpha_{\pm}=\left(\frac{\left(U\mp\sqrt{16t^{2}+U^{2}}\right)^{2}}{8t^{2}}+2\right)^{-1/2},
\end{equation}
\begin{eqnarray}
\beta_{\pm}=\frac{U\mp\sqrt{16t^{2}+U^{2}}}{4t}.
\end{eqnarray}

Therefore, they are no longer simple product states in the computational basis. As a result, the preparation of the eigenstates is no longer straightforward and we need to craft a procedure to initialize these states. We use the algorithm by Shende et al. \cite{shende06} to do so (see Fig.~\ref{fig:iniHUB}).

\subsection{Trotter-Suzuki formulas}

In general, the Hamiltonian of most interacting systems consists in a sum of non-commuting terms. This is the case of the Hubbard Hamiltonian in Eq.~(\ref{eq:hub2}), which is the sum of two terms that do {\em not} commute with each other. Since we would like to simulate the time propagator by exponentiating each Hamiltonian term individually (Appendix \ref{sec:controlled-U}) and multiplying them together, we can use the Trotter-Suzuki approximations. These allow us to express the exponential of a sum of non-commuting operators in terms of products of unitaries. To do so, each group of commuting terms is exponentiated individually for a small time-step and the exponentials are multiplied in such a way as to provide a reasonable approximation of the time-evolution operator. The first and second-order Trotter-Suzuki expansions are given by \cite{trotter1959product, suzuki1990fractal}
\begin{equation}
e^{\left(A+B\right)\tau}=\left(e^{A\tau/n}e^{B\tau/n}\right)^{n}+\xi_1,
\label{eq:1st_Trotter}
\end{equation}
\begin{equation}
e^{\left(A+B\right)\tau}=\left(e^{A\tau/2n}e^{B\tau/n}e^{A\tau/2n}\right)^{n}+\xi_2,
\label{eq:2nd_Trotter_0}
\end{equation}
where $\left[A,B\right]\neq0$, and we have errors $\xi_1=O(\tau\Delta \tau)$ and $\xi_2=O(\tau\left(\Delta \tau\right)^{2})$. Since $A$ commutes with itself, Eq.~(\ref{eq:2nd_Trotter_0}) can be recast into the product of fewer exponentials as
\begin{equation}
e^{\left(A+B\right)\tau}=e^{\frac{A\tau}{2n}}(e^{\frac{B\tau}{n}}e^{\frac{A\tau}{n}})^{n-1}e^{\frac{B\tau}{n}}e^{\frac{A\tau}{2n}}+\xi_2.
\label{eq:2nd_Trotter}
\end{equation}

The approximation is improved by increasing $n$, the {\em Trotter number}, thus making the time step $\Delta \tau = \tau/n$ small compared to the smallest time scale of $\mathcal{U}\left(\tau\right)$. As more repetitions are made, or equivalently as $\Delta \tau \rightarrow 0$, the error vanishes. However, there is a trade-off: making $\Delta \tau$ small, requires increasing the number of steps, and thereby the number of gates, which leads to errors when the algorithm is implemented in noisy hardware.

A good approximation is also possible if we take the evolution time to be much smaller than the smallest time scale of the operator. In the limit $\tau\rightarrow0$, the errors in Eq.~(\ref{eq:1st_Trotter}) and Eq.~(\ref{eq:2nd_Trotter}) scale asymptotically as
\begin{equation}
\xi_{1}\;\overset{\tau\rightarrow0}{\sim}\;\frac{\tau\Delta\tau}{2}\left[B,A\right],
\label{eq:Err_1st_Trotter}
\end{equation}
\begin{equation}
\xi_{2}\;\overset{\tau\rightarrow0}{\sim}\;\frac{\tau\left(\Delta\tau\right)^{2}}{24}\left(\left[A,\left[A,B\right]\right]+2\left[B,\left[A,B\right]\right]\right).
\label{eq:Err_2nd_Trotter}
\end{equation}

In QPE, we need to simulate $\mathcal{U}\left(\tau\right)$ raised to increasing powers of $2$. This is done by absorbing the exponent into $\tau$ and implementing $\mathcal{U}^{2^{k-1}}\left(\tau\right)=\mathcal{U}\left(\tau2^{k-1}\right)$, in which case we have $\xi_{1}=O(\tau\Delta \tau4^{k-1})$ and $\xi_{2}=O(\tau\left(\Delta \tau\right)^{2}8^{k-1})$, and the right-hand sides of Eq.~(\ref{eq:Err_1st_Trotter}) and Eq.~(\ref{eq:Err_2nd_Trotter}) are multiplied by $4^{k-1}$ and $8^{k-1}$, respectively. Due to this dependence of the error in the QPE exponent, we should let $n\rightarrow n2^{k-1}$ each time the propagator is exponentiated, both if we use the first- or the second-order decompositions. Increasing $n$ in this way, we cancel higher orders of the exponent in the error and guarantee it increases only linearly with the simulated exponent, achieving the required $\xi_{1}=O(\tau\Delta \tau2^{k-1})$ and $\xi_{2}=O(\tau\left(\Delta \tau\right)^{2}2^{k-1})$. Thus, circuit depth increases exponentially with $R$.

We are now in a position to discuss the quantum circuit that implements the time evolution operator of the Hubbard dimer. First, we shift the Hamiltonian in Eq.~(\ref{eq:hub2}) to drop the constant term, which leaves us with only three terms to exponentiate. The new eigenvalues are given in terms of the original ones in Eq.~(\ref{eq:hub2-eig}) by $\varepsilon{}_{i}'=\varepsilon{}_{i}-U/2$. Second, we separate the non-commuting terms of the Hamiltonian into two sets and implement each of these controlled propagators as in Fig.~\ref{fig:cU-Hub}. Applying these circuits with the right ordering and rotation parameters builds the desired Trotter order and number. We choose to perform our experiments by implementing the Trotter approximated propagator $\tilde{\mathcal{U}}\left(\tau\right)$ with the first order formula in Eq.~(\ref{eq:1st_Trotter}) and $n=1$, in the $\tau\rightarrow0$ limit.

\begin{figure}[h!]
\begin{flushright}
\begin{tabular}{ccc}
\centering{\
\Qcircuit @C=1em @R=1.15em {
   \lstick{} & \ctrl{2} & \ctrl{3} & \qw \\
   & & & & & \\
  \lstick{q_{1}: \ \ } & \gate{R_x(\theta_{1})} & \qw & \qw \\
   \lstick{q_{2}: \ \ } & \qw & \gate{R_x(\theta_{2})} & \qw
} \par} &  & \centering{\
\Qcircuit @C=1em @R=1.6em {
   \lstick{} & \qw & \ctrl{3} & \qw & \qw \\
   & & & & & \\
   \lstick{} & \ctrl{1} & \qw & \ctrl{1} & \qw \\
  \lstick{} & \targ & \gate{R_z(\theta_{3})} & \targ & \qw
} \par}\tabularnewline
 &  & \tabularnewline
$\left(a\right)$ &  & $\left(b\right)$\tabularnewline
\end{tabular}
\par\end{flushright}
\caption{Controlled implementation of the propagators coming from the two groups of commuting terms in Eq.~(\ref{eq:hub2}). Using the first order approximation, we have $\theta_{1}=\theta_{2}=-2t\tau/n$ and $\theta_{3}=U\tau/n$, while with Eq.~(\ref{eq:2nd_Trotter}), $\theta_{3}$ remains the same but $\theta_{1}=\theta_{2}$ are divided by $2$ for the first and last exponentials. As explained, the exponents of the phase estimation algorithms are introduced as $2^{k-1}$ factors in $\theta_1$, $\theta_2$ and $\theta_3$.}
\label{fig:cU-Hub}
\end{figure}
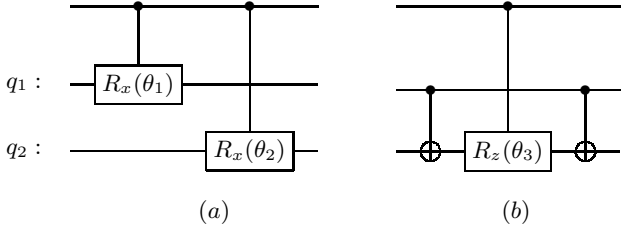

\begin{figure}[h]
\begin{centering}
\begin{tabular}{ccc}
\centering{\
\Qcircuit @C=0.5em @R=1em {
   \lstick{} & \ctrl{1} & \qw \\
   \lstick{} & \gate{R_x(\theta)} & \qw } \par} & $\begin{array}{c}
\\
\\
=
\end{array}$ & \centering{\
\Qcircuit @C=0.3em @R=1em {
   \lstick{} & \qw & \ctrl{1} & \qw & \ctrl{1} & \qw & \qw \\
   \lstick{} & \gate{U_1(\frac{\pi}{2})} & \targ & \gate{U_3(-\frac{\theta}{2},0,0)} & \targ & \gate{U_3(\frac{\theta}{2},-\frac{\pi}{2},0)} & \qw } \par}\tabularnewline
\end{tabular}
\par\end{centering}
\caption{Implementation of the controlled-$R_{x}\left(\theta\right)$ operation using IBM's platform gate set in Eq.~(\ref{eq:ibmgateset}).}
\label{fig:cRx}
\end{figure}
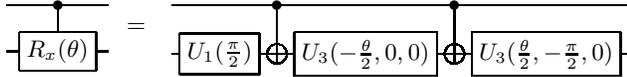

\section{Results for the Hubbard dimer \label{sec:hubres}}

\begin{figure}
\noindent \begin{centering}
\centering{\
\Qcircuit @C=1em @R=1em {
\lstick{q_{1}:} & \qw & \targ & \gate{R_y\left(\theta\right)} & \targ & \gate{R_y\left(\pi/2\right)} & \qw \\
  \lstick{q_{2}:} & \gate{R_y\left(\pi/2\right)} & \ctrl{-1} & \qw & \ctrl{-1} & \qw & \qw
} \par}
\par\end{centering}
\caption{Initialization circuit for $\left|\varepsilon_{-}\right\rangle$. The exact amplitudes of the groundstate obtained with $t=0.35$ and $U=0.2$ are reproduced with $\theta=0.14189705(...)$, input with the maximum possible numerical precision by the IBM Q platform.}
\label{fig:iniHUB}
\end{figure}
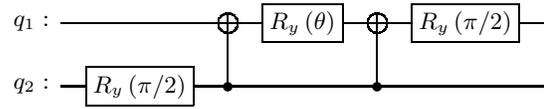

Let us now discuss the QPE results for the DQS of the Hubbard dimer at half-filling. For the purposes of testing our methodology, we focus on the groundstate of the model and choose the Hamiltonian parameters in Eq.~(\ref{eq:hub2}) to be $t=0.35$ and $U=0.2$. The procedure is identical for different eigenstates and parameter choices. The theoretically expected eigenvalue is thus $\varepsilon_{-}\approx-0.6071$. To prepare this state, we use the circuit in Fig.~\ref{fig:iniHUB} to bring simulation register from $\ket{00}$ to $\ket{\varepsilon_{-}}$.

Due to the stringent circuit depth requirements, we use the IPEA to reduce the noise level in the final computation as much as possible. We choose to carry out the experiments with the non-exhaustive version of the algorithm for two reasons. The first one is logistical: currently, IBM Q does not allow for measurement feedback and state re-setting of the quantum circuits. This makes performing the exhaustive IPEA a lengthy task when $R$ is sufficiently large because of the need to request a high number of circuit executions to the system and going through a communication bottleneck. The second reason is methodological: when Trotterization is needed, we cannot avoid working with a superposition state if we have a fixed initialization procedure across different values of $\tau$. This is because the Trotter approximated propagator, $\tilde{\mathcal{U}}\left(\tau\right)$, does not commute with itself at different times and there is no common basis for the operator for all $\tau$. Therefore, if we just want to probe the phase coming from one of the eigenstates (the groundstate in our case) we cannot use the mean phase direction estimator to do so directly and without further post-processing, because the $0.\phi$ frequency counts will include the contribution of nearby eigenstates. The use of the mean phase direction estimator with superposition states is not explored here.

We must now remember that, apart from quantum localization effects that can bound the Trotter error periodically in $\tau$ even at long evolution times \cite{heyl2019quantum}, the approximation is only valid for $\tau\approx0$ if a small Trotter $n$ is to be used. Hence only short evolution times should be simulated for the obtained phase estimates to be adequate for regression with a linear function $f(\tau) = m\tau + b$ in Eq.~(\ref{eq:X2circ}). Since we want to maximize the number of sampled $\tau$ points to improve the quality of the regression, we also simulate negative time to obtain a symmetric time interval close to, and centered at, $\tau=0$. We perform experiments for 200 points in $\left|\tau\right|<5$.

\begin{figure}
\includegraphics[width=1.\linewidth]{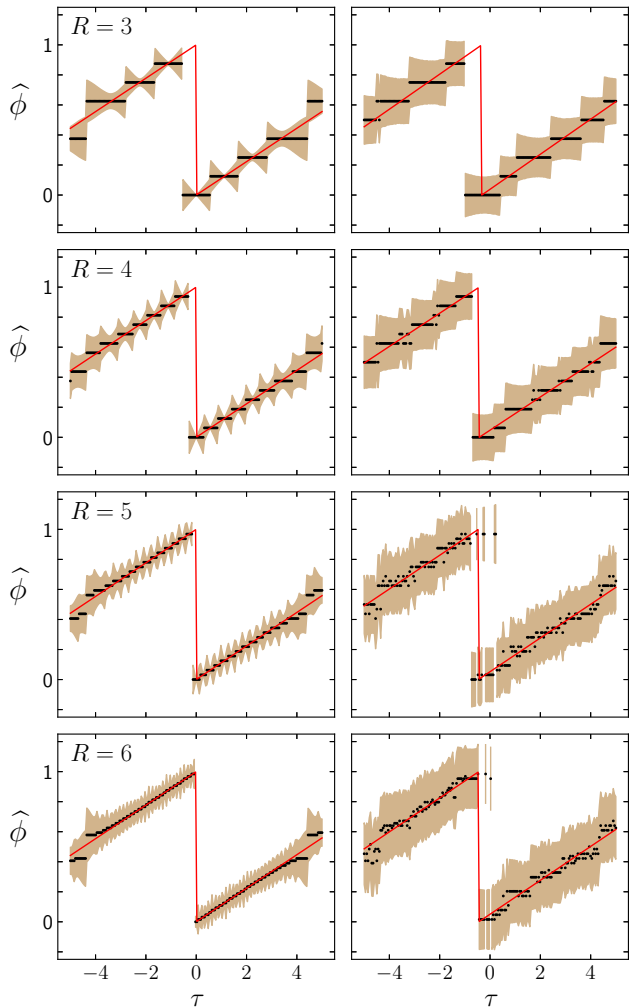}
\caption{Unitary simulator (left) and experimental (right) results for the non-exhaustive IPEA with the Hubbard model, using a $1^{st}$ order Trotter approximation with $n=1$, and varying resolution from 3 to 6. The fitted line is obtained with the points for $\left|\tau\right|<3$ (see Tables~\ref{tab:Hub-exp} and \ref{tab:Hub-un}).}
\label{fig:HUB1}
\end{figure}

\begin{figure}
\includegraphics[width=1.\linewidth]{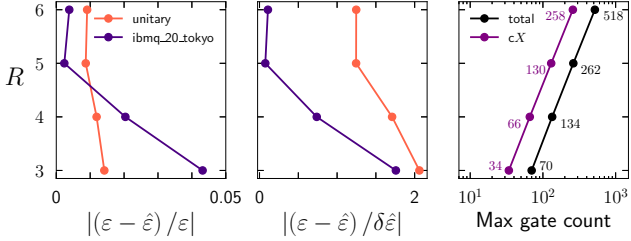}
\caption{Relative errors, absolute standard score and number of gates in the most demanding iteration for each non-exhaustive IPEA experiment with the Hubbard dimer. Both the experimental results and those obtained in a unitary classical simulator are shown as a function of the resolution $R$.}
\label{fig:HUB2}
\end{figure}

In Fig.~\ref{fig:HUB1} we show the experimental results for 4 levels of resolution, as well as the same computational results obtained in a simulated fault-tolerant quantum computer, for comparison. For each timestamp, $\widehat{\phi}\left(\tau\right)$ is plotted with the standard deviation of the reconstructed LPMF represented by the brown region. Every point is the result of $R$ iterations, each one with $O=5000$ observations. As in the case of the computations in section~\ref{sec:spinres}, experimental dispersion on the right-hand side column is much larger than what would be expected for ideal quantum computations, deteriorating with increasing $R$ due to the higher vulnerability to noise and decoherence that comes with the increase in circuit depth. Nonetheless, as we describe below, we find a good agreement between theory and experiment when we fit the experimental datapoints to the non-noisy theoretical model. We could not increase $R$ further because with $R=6$ we already achieved the maximum number of gates that IBM allows in real hardware (see Fig.~\ref{fig:HUB2}).

To understand these results, one should naturally expect that the initial prepared state on the simulation register, while being an eigenstate of the Hubbard Hamiltonian, is not an eigenstate of $\tilde{\mathcal{U}}\left(\tau\right)$, on account of the Trotter approximation. This in turn means that the initial state has a decomposition in the eigenstates of $\tilde{\mathcal{U}}\left(\tau\right)$. For short evolution times, the overlap of the initial state with the groundstate of $\tilde{\mathcal{U}}\left(\tau\right)$ is large, however, as $\tau$ increases, the initial state progressively departs from this eigenket and eventually starts to overlap the most with some other nearby state, causing the maximum peak in the probability function to suddenly jump across basis states, as can be seen around $\tau=4$.

We choose a smaller interval to fit only the datapoints $\left|\tau\right|<\tau_{\max}$, with $\tau_{\max}=3$. The fitting parameters are displayed in Table~\ref{tab:Hub-exp}, Table~\ref{tab:Hub-un} and Fig.~\ref{fig:HUB2}, and the fitted model is plotted in red in Fig.~\ref{fig:HUB1}. We can see that the most accurate results obtained with \emph{ibmq\_20\_tokyo} are quite good, with a relative error of $\approx0.2\%$ for $R=5$ and $\approx0.4\%$ for $R=6$. Only these two experiments are more accurate in \emph{ibmq\_20\_tokyo} than in unitary simulations. This is interpreted as a fortuity statistical occurrence which is actually not significant since the precision error $\delta\hat{\varepsilon}$ of the estimation is consistently smaller in the unitary experiments.

Importantly, unitary simulations show we hit an asymptotic limit where the accuracy error is not tending towards 0, because in the chosen $|\tau|<\tau_{\max}$ interval, the datapoints do not trace a perfectly straight line due to the inaccuracy of the first order Trotter approximation.

This brings us to the most important conclusion learnt from this methodology. Increasing accuracy while reducing the standard score by linearly fitting $\widehat{\phi}\left(\tau\right)$ requires using the $\tilde{\mathcal{U}}\left(\tau\right)$ decomposition that best matches the actual time propagator of the system in the range of simulated $\tau$ values. This can be done by either increasing the Trotter order while keeping a fixed $\tau_{\max}$, or by fixing a Trotter order and a Trotter number (in our case, first order and $n=1$) and limiting regression to shorter evolution times, that is zooming into $\tau=0$ by reducing $\tau_{\max}$. The challenge posed by the second option is that we start to hit the $0.\phi$ resolution limit for the phase readouts to reliably fit the data, calling for the necessity to increase $R$ simultaneously with the decrease in $\tau_{\max}$ to avoid obtaining a stepwise signal.

Thus, if increasing $R$ increases the number of gates exponentially, it is also true that it increases the number of readout basis states $0.\phi$ exponentially. This allows us to perform an exponential zoom into $\tau=0$ and get an exponential increase in accuracy, since we beat the scaling of the Trotter error, which is polynomial. The ability to achieve chemical accuracy thus depends on the ability to implement exponentially deeper circuits. The procedure would be to fix initial values for $\tau_{\max}$ and $R$ to perform the first experiment and then, each time $R$ is increased by one unit, $\tau_{\max}$ would be divided by 2.

\begin{table}[ht]
\newcolumntype{R}{>{\centering\arraybackslash}X}
\noindent
\begin{tabularx}{\columnwidth}{ *{5}{R} }
\hline
\hline
$R$ & $3$ & $4$ & $5$ & $6$ \smallskip{}\\
\hline
\multirow{2}{*}{$m$} & $0.1111$ & $0.1114$ & $0.1117$ & $0.1117$\\
 & $\pm 0.0007$\smallskip{}
 & $\pm 0.0007$ & $\pm 0.0007$ & $\pm 0.0007$\\
\multirow{2}{*}{$b$} & $0.0000$ & $0.0000$ & $0.0002$ & $0.0001$\\
 & $\pm 0.0004$\smallskip{}
 & $\pm 0.0008$ & $\pm 0.0004$ & $\pm 0.0005$\\
$\chi_{\text{circ}}^2/\text{ndf}$ & $1.63$\smallskip{}
 & $0.38$ & $0.10$ & $0.03$\\
\multirow{2}{*}{$\hat{\varepsilon}$} & $-0.599$ & $-0.600$ & $-0.602$ & $-0.602$\\
 & $\pm 0.004$\smallskip{}
 & $\pm 0.005$ & $\pm 0.004$ & $\pm 0.004$\\
$\left(\varepsilon-\hat{\varepsilon}\right)/\left|\varepsilon\right|$ & $-0.013$\smallskip{}
 & $-0.012$ & $-0.008$ & $-0.008$\\
$\left(\varepsilon-\hat{\varepsilon}\right)/\delta\hat{\varepsilon}$ & $-2.03$ & $-1.42$ & $-1.28$ & $-1.28$\\
\hline
\hline
\end{tabularx}
\caption{Fitting parameters for the unitary results of the non-exhaustive IPEA with the Hubbard dimer.}
\label{tab:Hub-un}
\end{table}

\begin{table}[ht]
\newcolumntype{R}{>{\centering\arraybackslash}X}
\noindent
\begin{tabularx}{\columnwidth}{ *{5}{R} }
\hline
\hline
 $R$ & $3$ & $4$ & $5$ & $6$ \smallskip{}\\
\hline
\multirow{2}{*}{$m$} & $0.1167$ & $0.1106$ & $0.1123$ & $0.1129$\\
 & $\pm 0.0024$\smallskip{}
 & $\pm 0.0027$ & $\pm 0.0031$ & $\pm 0.0034$\\
\multirow{2}{*}{$b$} & $0.0382$ & $0.0473$ & $0.0522$ & $0.0504$\\
 & $\pm 0.0040$\smallskip{}
 & $\pm 0.0046$ & $\pm 0.0053$ & $\pm 0.0059$\\
$\chi_{\text{circ}}^2/\text{ndf}$ & $0.42$\smallskip{}
 & $0.17$ & $0.18$ & $0.12$\\
\multirow{2}{*}{$\hat{\varepsilon}$} & $-0.633$ & $-0.595$ & $-0.606$ & $-0.609$\\
 & $\pm 0.015$\smallskip{}
 & $\pm 0.017$ & $\pm 0.019$ & $\pm 0.022$\\
$\left(\varepsilon-\hat{\varepsilon}\right)/\left|\varepsilon\right|$ & $0.043$\smallskip{}
 & $-0.020$ & $-0.002$ & $0.004$\\
$\left(\varepsilon-\hat{\varepsilon}\right)/\delta\hat{\varepsilon}$ & $1.76$ & $-0.74$ & $-0.07$ & $0.11$\\
\hline
\hline
\end{tabularx}
\caption{Fitting parameters for the experimental results of the non-exhaustive IPEA with the Hubbard dimer.}
\label{tab:Hub-exp}
\end{table}

\section{Discussion and Conclusions\label{dc}}

In this work, we have looked at the use of QPE algorithms for the determination of Hamiltonian eigenvalues both from a theoretical and an experimental points of view. We explored the scenario of eigenstate initialization of these algorithms, setting the basis for more general treatments using our methods.

On the first front, we have derived an expression for the first trigonometric moment about the mean direction of the fault-tolerant parent probability distribution associated with eigenstate inputs to full-blown QPE (the PEA and the exhaustive IPEA). This quantity can be inexpensively estimated from a sample, without bias, introducing a new avenue for classically post-processing the quantum computational results. We showed that using the sample mean phase direction as a stand-alone estimator of the phase carries a lower accuracy error bound than using the standard majority rule. Moreover, it can be further post-processed to construct an unbiased estimator of the phase by inverting Eq.~(\ref{eq:mpd}), in this initial state scenario. Despite working in the scenario where the input state is an eigenstate, the mean phase direction is also the basis for a generalization to post-processing superposition state inputs, a project we leave as future research.

We then presented the use of this newly-introduced quantity to directly find the energy levels of Hamiltonians that do not require Trotterized simulation. With this approach, we bypass the need to increase the resolution in the quantum algorithm and trade it by sampling and post-processing, avoiding the limitations associated with higher circuit depth in NISQ devices. To simulate Hamiltonians with non-commuting terms, we explored the use of the first-order Trotter-Suzuki decomposition with majority-rule QPE. The contribution of the mean phase direction for the post-processing of this scenario also requires further exploration.

On the experimental side of this project, we provided proof-of-concept demonstrations of our methods using the quantum computing devices from IBM Q. We tested this DQS program using three simple model Hamiltonians for which the quantum circuits were provided. Two of these models were studied with the PEA and our proposed estimator, while the third one was simulated using the IPEA with the standard majority rule. To characterize dispersion of the results obtained with the non-exhaustive IPEA, we proposed a way to reconstruct a lossy probability distribution over all the possible $R$-bit sequences by using the two-state histograms obtained at each iteration; this approach was used to calculate the phase circular standard deviation.

Experiments showed that noise can degrade the fault-tolerant probability distributions enough to bias the estimators sometimes. Because of that, we had a second reason to execute the programs for several timesteps while varying a control parameter $\tau$ in the time propagator of the simulated systems. This allowed reducing variability by fitting the experimental data to the expected theoretical model to estimate the energy levels through the fitting parameters. As can be seen from the final results, this method proved its robustness. If necessary, the precision of the final energy level estimations of the systems implemented with non-trotterized propagators could in principle be further improved by simulating and fitting longer evolution times.

Despite demonstrating our methodology with such simple models, our methods are fully general and can be applied to more complex hamiltonians as soon as the hardware develops enough to permit so. Thanks to the exhaustive version of the IPEA, it becomes apparent that the main limitation for QPE-based digital quantum simulation on noisy near-term devices rests on the implementation the propagator of the system. Further progress requires the availability of quantum hardware with a balanced improvement of the number of qubits, connectivity, coherence time and error rates, such that it becomes possible to reliably execute wider and deeper circuits.

\ \

{\it Acknowledgements}. We thank the IBM corporation for making quantum computers available for the community. R. G. acknowledges the INL summer student program. 
G.C. acknowledges Funda\c{c}\~{a}o para a Ci\^{e}ncia e a Tecnologia (FCT) for Grant No. SFRH/BD/138806/2018.
J.F.R. and G.C. acknowledge the FCT for grant PTDC/FIS-NAN/4662/2014 (016656). We thank Bruno Murta for fruitful discussions.

\ \

\appendix

\section{QPE with superposition states \label{sec:QPE-super}}

Here we discuss the workings of the PEA \cite{Nielsen,cleve98} when the initial state $\ket{\psi}$ is not an eigenstate of ${\cal U}$. We write
\begin{equation}
\ket{\psi}=\sum_n c_n|\phi_n\rangle,
\end{equation}
where $\sum_n |c_n|^2=1$, and
\begin{equation}
{\cal U}\ket{\psi}= \sum_n c_n e^{2\pi i \phi_n} \ket{\phi_n}.
\end{equation}

After the controlled-${\cal U}$ operations, the quantum state of the computer can be written up as
\begin{equation}
\ket{\Psi_{II}} = \sum_{n} c_n \left( \bigotimes_{k=1}^{R} \frac{\ket{0} + e^{i 2\pi 2^{k-1} \phi_n} \ket{1}}{\sqrt{2}} \right) \otimes \ket{\phi_n}.
\end{equation}

The permuted inverse quantum Fourier Transform maps the state of the phase register qubits that is associated with each $\ket{\phi_n}$ into a state of the computational basis given by
\begin{widetext}
\begin{align}
\ket{\Psi_{III}^{n}}_{\mathrm{PR}} & =QFT^{\dagger}\left[\bigotimes_{k=1}^{R}\frac{\ket{0}+e^{i2\pi2^{k-1}\phi_{n}}\ket{1}}{\sqrt{2}}\right]\nonumber \\
& =\sum_{b_{1,n}=0}^{1}\sum_{b_{2,n}=0}^{1}\cdots\sum_{b_{R,n}=0}^{1}\left(\sum_{k=1}^{2^{R}}\frac{e^{i2\pi\left(\phi_{n}-0.b_{1,n}b_{2,n}\cdots b_{R,n}\right)\left(k-1\right)}}{2^{R}}\right)\ket{b_{1,n}}\ket{b_{2,n}}\cdots\ket{b_{R,n}},
\end{align}
\end{widetext}
where we have expressed the phases as binary fractions
\begin{equation}
0.\phi_n= 0.b_{1,n}b_{2,n}\cdots b_{R,n} = \sum_{k=1}^{R} \frac{b_{k,n}}{2^{k}}.
\end{equation}

When $\phi_n$ can be expressed exactly in $R$ bits, this returns a single basis state given by
\begin{equation}
\ket{\Psi_{III}^{n}}_{\mathrm{PR}} = \bigotimes_{k=1}^{R}\ket{b_{k,n}}=\ket{0.\phi_n},
\end{equation}
such that we get the phase $\phi_n$ with probability $|c_n|^2$. When it cannot, the outcome of the PEA before the projective readout on the phase register is 
\begin{equation}
\ket{\Psi_{III}} = \sum_{n} c_n \ket{\Psi_{III}^{n}}_{\mathrm{PR}} \otimes \ket{\phi_n}.
\end{equation}
Thus, after measurement, we obtain the parent distribution given by Eq.~(\ref{eq:PMF_PEA}) with probability $|c_n|^2$.
As a consequence, a given $0.\phi$ readout, e.g. 110 for the case of $R=3$, can correspond to the collapse of different orthogonal states of the simulation register. 
In this way, the readout produces a distillation of the $\ket{\phi_n}$ states that is no longer a linear superposition.

\section{Implementation of controlled-$\mathcal{U}$ operations in terms of elementary gates\label{sec:controlled-U}}

The name of the game in phase estimation is to implement the time evolution operator $\mathcal{U}\left(t\right)=e^{-it\mathcal{H}/\hbar}$. That can be done once we have the Hamiltonian translated into the qubit language as a sum of terms $\mathcal{H}=\sum_{m}h_{m}$, each one with a tensor product structure of $X$, $Y$, $Z$ Pauli operators. Exponentiation of each $h_{m}$ term is easily carried out as follows \cite{Nielsen}.

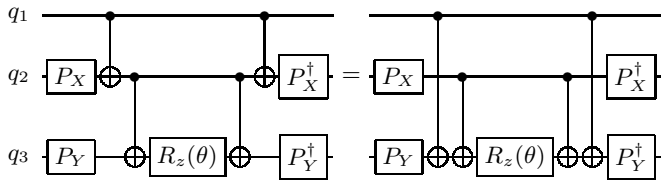
\begin{figure}[hbt]
\begin{raggedleft}
\begin{tabular}{c}
\Qcircuit @C=0.19em @R=1.5em {
   \lstick{q_{1}} & \qw & \ctrl{1} & \qw & \qw & \qw & \ctrl{1} & \qw & \qw & \; & \; & \; & \; & \; & \; & \; & \; & \qw & \ctrl{2} & \qw & \qw & \qw & \ctrl{2} & \qw & \qw  \\
   \lstick{q_{2}} & \gate{P_{X}} & \targ & \ctrl{1} & \qw & \ctrl{1} & \targ & \gate{P_{X}^{\dagger}} & \qw & \; & \; & \;  & = & \; & \; & \; & \; & \gate{P_{X}} & \qw & \ctrl{1} & \qw & \ctrl{1} & \qw & \gate{P_{X}^{\dagger}} & \qw\\
   \lstick{q_{3}} & \gate{P_{Y}} & \qw & \targ & \gate{R_z(\theta)} & \targ & \qw & \gate{P_{Y}^{\dagger}} & \qw & \; & \; & \; & \; & \; & \; & \; & \; & \gate{P_{Y}} & \targ & \targ & \gate{R_z(\theta)} & \targ & \targ & \gate{P_{Y}^{\dagger}} & \qw}\\
\end{tabular}
\par\end{raggedleft}
\caption{Implementation of a general Hamiltonian term exemplified with $\exp\left(-i \theta Z_{1}X_{2}Y_{3}/2\right)$. Two equivalent ways of computing parity are shown.}
\label{fig-general_cU}
\end{figure}

First recall that $R_{z}\left(\theta\right)=\exp{(-i\theta Z/2)}$ acts on a general qubit state by adding a phase $\exp{(-i\theta/2)}$ (resp. $\exp{(i\theta/2)}$) to the $\left|0\right\rangle $ (resp. $\left|1\right\rangle $) basis state. Now take an operator of the form $\exp{(-i\theta Z_{1}Z_{2}\cdots Z_{n}/2)}$. Its effect is to apply the phase $\exp{(-i\theta/2)}$ (resp. $\exp{(i\theta/2)}$) to the computational basis states with an even (resp. odd) number of qubits in state $\left|1\right\rangle$. To reproduce this operation with a $n$-qubit quantum circuit, the approach lies in entangling all the qubits with a cascade of $n-1$ $\mathrm{c}X$ gates targeting an arbitrarily chosen \emph{parity qubit} where a single $R_{z}\left(\theta\right)$ gate acts before the same $\mathrm{c}X$ cascade is applied again. The task of the first batch of $\mathrm{c}X$ gates is to ensure the parity qubit is in the $\left|0\right\rangle$ (resp. $\left|1\right\rangle$) state if parity is even (resp. odd) before the $R_{z}\left(\theta\right)$ gate is applied to it, while the second cascade returns this phase back to the original basis state. Two different ways of constructing the $\mathrm{c}X$ cascade are shown in Fig.~\ref{fig-general_cU}. The advantage of each approach depends on the coupling architecture of the computer. Mixing both approaches is also possible.

If the exponential contains $X$ or $Y$ matrices, we need to change to their diagonal basis in between the full circuit, that is $X=P_{X}^{\dagger}ZP_{X}$ and $Y=P_{Y}^{\dagger}ZP_{Y}$ with $P_{X}=H$ and $P_{Y}=R_{x}\left(\pi/2\right)$. Thus, it suffices to apply a $P_{X}$ and $P_{Y}$ gate on the correct qubits before computing parity, and their adjoint at the end of the circuit, as also exemplified in Fig.~(\ref{fig-general_cU}). To make it a controlled operation, only the action of the $Z$-rotation gate needs to be controlled, as all the other operators are applied together with their adjoints, yielding the identity when the control qubit is set to $\left|0\right\rangle$.

\bibliography{biblio}{}

\end{document}